\documentclass[12pt]{article}
\usepackage{bbm}
\usepackage{color}
\usepackage{authblk}
\usepackage{latexsym}
\usepackage{epsfig,amssymb,euscript, mathrsfs,cite}
\usepackage{amsmath}
\usepackage{mathrsfs} 
\usepackage{enumerate}
\definecolor{MyDarkBlue}{rgb}{0.15,0.15,0.45}
\usepackage[linktocpage=true]{hyperref}
\hypersetup{
colorlinks=true,
citecolor=MyDarkBlue,
linkcolor=MyDarkBlue,
urlcolor=MyDarkBlue,
pdfauthor={},
pdftitle={},
pdfsubject={hep-th}
}
\newsavebox{\ns}
\newsavebox{\dbrane}
\newsavebox{\dbshort}

\def\be{\begin{equation}}
\def\ee{\end{equation}}
\def\bea{\begin{eqnarray}}
\def\eea{\end{eqnarray}}

\newcommand{\nn}{\nonumber\\}

\newcommand\R{\mathbb{R}}
\newcommand\Z{\mathbb{Z}}

\newcommand\C{\mathbb{C}}

\newcommand\diff{\mathrm{d}}

\newcommand{\ii}{\mathrm{i}}

\newcommand{\ex}{\mathrm{e}}
\newcommand{\vol}{\mathrm{vol}}
\newcommand{\Vol}{\mathrm{Vol}}

\newcommand{\p}{p}

\newlength{\sswidth}

\numberwithin{equation}{section}       

\newcommand{\CC}{\kappa}
\newcommand{\UJ}{U(1)_{\mathcal{J}}}
\newcommand{\AJ}{\mathcal{A}_{\mathcal{J}}}
\newcommand{\FJ}{\mathcal{F}_{\mathcal{J}}}
\newcommand{\JJ}{\mathcal{J}}

\begin{document}

\begin{titlepage}

\vskip 1cm

\begin{center}


{\Large \bf Twisted D3-brane and M5-brane compactifications}

\vskip 0.5cm

{\Large \bf  from multi-charge spindles}

\vskip 2.0cm

{Andrea Boido,
 Juan Manuel P\'erez Ipi\~na and James Sparks}

\vskip 0.5cm

\textit{Mathematical Institute, University of Oxford,\\
Andrew Wiles Building, Radcliffe Observatory Quarter,\\
Woodstock Road, Oxford, OX2 6GG, U.K.\\}

\end{center}

\vskip 2.0 cm

\begin{abstract}
\noindent  We construct families of supersymmetric AdS$_3\times Y_7$ 
and AdS$_3\times Y_8$ solutions to type IIB string 
theory and M-theory, respectively. Here $Y_7$ is an $S^5$ fibration over $\Sigma$, while $Y_8$ is an $S^4$ fibration over 
$\Sigma_g\times \Sigma$, where $\Sigma_g$ is a  Riemann surface of genus $g>1$ and $\Sigma$ is a 
two-dimensional orbifold  known as a spindle. 
We interpret the solutions
as near-horizon limits of $N$ D3-branes wrapped on 
$\Sigma$ and $N$ M5-branes wrapped on $\Sigma_g\times \Sigma$, respectively. 
These
are holographically dual to $d=2$, $(0,2)$ SCFTs, 
and we show that the central charge and superconformal R-symmetry of 
the gravity solutions agree with dual field theory calculations.
\end{abstract}

\end{titlepage}

\pagestyle{plain}
\setcounter{page}{1}
\newcounter{bean}
\baselineskip18pt

\tableofcontents

\newpage

\section{Introduction and summary}\label{sec:intro}

Interesting classes of superconformal field theories (SCFTs) may be 
obtained in string theory and M-theory by wrapping branes on a 
compact space $\Sigma$, and flowing to the infrared (IR).  
Supergravity solutions describing the near-horizon limits 
of such wrapped branes were first constructed in the 
pioneering work of \cite{Maldacena:2000mw}, describing 
D3-branes and M5-branes wrapping a constant curvature 
Riemann surface $\Sigma_g$ of genus $g>1$. This work has  subsequently been
generalized in many directions, leading to a large literature on the subject.  

Until recently such constructions have realized 
supersymmetry on the wrapped brane worldvolume via a so-called topological 
twist \cite{Witten:1988ze}. Here the spin connection on the wrapped directions
$\Sigma$ is effectively cancelled by the coupling of the spinor to external
R-symmetry gauge fields, which geometrically are connections 
on the normal bundle to $\Sigma$ in spacetime. 
The upshot is that the preserved Killing spinors are sections 
of trivial bundles over $\Sigma$, and in fact constant. 
A different realization of supersymmetry has 
 recently been presented in \cite{Ferrero:2020laf}. 
Here a family of supersymmetric AdS$_3\times \Sigma$ 
solutions was  constructed in $d=5$ minimal gauged supergravity, 
where $\Sigma$ is a two-dimensional orbifold surface known 
as a spindle. This is topologically a two-sphere, but with conical 
deficit angles $2\pi (1-\frac{1}{n_\pm})$ at the two poles, 
specified by coprime positive integers $n_+\neq n_-$. 
Equivalently, $\Sigma=\mathbb{WCP}^1_{[n_-,n_+]}$ is 
a weighted projective space. When $n_--n_+$ is divisible by 
3 these may be uplifted on an $S^5$ internal space, 
leading to AdS$_3\times Y_7$ solutions of type IIB supergravity in 
which $Y_7$ takes the fibred form
\begin{align}\label{introfibre}
S^5 \ \hookrightarrow\  Y_7 \  \rightarrow\  \Sigma \, = \, \mathbb{WCP}^1_{[n_-,n_+]}\, .
\end{align}
Since these ten-dimensional solutions are sourced only by five-form flux $F_5$, 
with $N$ units of flux through the $S^5$ fibre of \eqref{introfibre}, 
these are naturally interpreted as near-horizon limits of $N$ D3-branes 
wrapped on the spindle $\Sigma$. However, the Killing spinors 
on $\Sigma$ are not constant, and moreover are sections of non-trivial 
spinor bundles \cite{Ferrero:2020twa}. The flux through $\Sigma$ of the $d=5$ Abelian R-symmetry 
gauge field in gauged supergravity is correspondingly not
simply proportional to the Euler number $\chi(\Sigma)$, given by 
\begin{align}\label{introEuler}
\chi(\Sigma)\, = \, \frac{n_-+n_+}{n_-n_+}\, ,
\end{align}
as is characteristic of a topological twist.

In the first part of 
this paper we generalize the construction of \cite{Ferrero:2020laf} 
by allowing for more general fibrations in \eqref{introfibre}. 
More precisely, the $d=5$ minimal gauged supergravity solution
of \cite{Ferrero:2020laf}  is  a special case of a more
general class of (local) AdS$_3$ solutions to $d=5$, $U(1)^3$ gauged supergravity 
with two additional parameters, first constructed in \cite{Kunduri:2007qy}. 
These also uplift on $S^5$, with the three gauge fields of $U(1)^3$ determining 
the fibration of this internal space over the $d=5$ spacetime. 
The resulting supersymmetric AdS$_3\times Y_7$ solutions were in fact
constructed before reference \cite{Kunduri:2007qy}, in 
\cite{Gauntlett:2006ns}. The approach we take to the global 
analysis of these supergravity solutions is different to that in 
\cite{Kunduri:2007qy}, \cite{Gauntlett:2006ns}, and instead follows 
 \cite{Ferrero:2020laf}. Specifically, we first construct 
 AdS$_3\times \Sigma$ solutions 
of the $d=5$, $U(1)^3$ gauged supergravity theory, where 
$\Sigma=\mathbb{WCP}^1_{[n_-,n_+]}$ is a spindle specified by coprime positive integers $n_->n_+$. 
We then appropriately quantize the three Abelian R-symmetry gauge field fluxes 
through $\Sigma$, in such a way that the resulting fibration \eqref{introfibre} 
is well-defined, with a smooth total space $Y_7$. More precisely, 
defining the fluxes
\begin{align}\label{introfluxes}
Q_I\, \equiv\, \frac{1}{2\pi}\int_\Sigma F^{(I)}\, ,
\end{align}
where $F^{(I)}$, $I=1,2,3$, are the three Abelian R-symmetry gauge field curvatures, 
we construct solutions with 
\begin{align}\label{introcharges}
Q_1\, = \, \frac{q}{n_-n_+}\, , \qquad Q_2\, = \, \frac{p}{n_-n_+}\, , \qquad Q_3\, = \, \frac{n_--n_+-q-p}{n_-n_+}\, .
\end{align}
Here $q,p>0$ are positive integers, constrained so that $q+p<n_--n_+$, and where $q$ and $p$ have no common factor 
with either of $n_-,n_+$. Notice that the charges \eqref{introcharges} satisfy
\begin{align}\label{introchargesum}
Q_1+Q_2+Q_3\, =\, \frac{n_--n_+}{n_-n_+}\, .
\end{align}
This should be contrasted with a topological twist, where instead on the right hand side of 
\eqref{introchargesum} one would have the Euler number of the spindle \eqref{introEuler}.
The solutions of \cite{Ferrero:2020laf} are recovered by setting 
$Q_1=Q_2=Q_3$, or equivalently $p=q=\frac{1}{3}(n_--n_+)$, which then
requires $n_--n_+$ to be divisible by 3. 

The central charge of these solutions is given by 
\begin{equation} \label{introc}
	c \, =  \, \frac{3 p q (n_- - n_+ - q - p)}{n_- n_+ [(n_--n_+) (p+q) + n_-n_+-p^2-p q-q^2]} N^2 \,.
\end{equation}
The solution is thus specified by five positive integers: $n_-, n_+$ determine the topology 
of the wrapped spindle $\Sigma=\mathbb{WCP}^1_{[n_-,n_+]}$, $q,p$ determine 
the R-symmetry fluxes \eqref{introcharges} and hence twisting in the fibration \eqref{introfibre}, 
and finally $N$ is the number of wrapped D3-branes.
As in  \cite{Ferrero:2020laf}, the description of these solutions suggests that 
the dual $d=2$, $(0,2)$ SCFTs arise from a twisted compactification 
of $\mathcal{N}=4$ SYM on $\Sigma$, where there are background 
R-symmetry fluxes given by \eqref{introcharges}. We make a
check on this conjecture by computing both the central charge and 
superconformal R-symmetry of the $d=2$ field theories, 
making use of the anomaly polynomial of $\mathcal{N}=4$ SYM 
together with $c$-extremization \cite{Benini:2012cz}. We find 
precise agreement with the corresponding supergravity quantities.

When $Q_1=Q_2$, or equivalently $p=q$, the solutions to
$d=5$, $U(1)^3$ supergravity are also solutions 
to  Romans  $\mathcal{N}=4$, $SU(2)\times U(1)$ supergravity \cite{Romans:1985ps}. The latter 
is a consistent truncation of $d=11$ supergravity on the Maldacena-N\'u\~nez solution 
referred to at the start of this introduction \cite{Gauntlett:2007sm}. We thus also obtain AdS$_3\times Y_8$
solutions of $d=11$ supergravity, where $Y_8$ takes the fibred form 
\begin{align}\label{introfibreM5}
S^4 \ \hookrightarrow\  Y_8 \  \rightarrow\  \Sigma_g\times \Sigma\, .
\end{align}
Here $\Sigma_g$ is a constant curvature Riemann surface of genus $g>1$.
The original Maldacena-N\'u\~nez solution corresponds 
to the AdS$_5$ vacuum of the Romans theory, and is dual to 
the $d=4$, $\mathcal{N}=2$ SCFT living on $N$ M5-branes wrapped on the 
Riemann surface $\Sigma_g$, which is holomorphically embedded inside a Calabi-Yau two-fold.  
The uplift to $d=11$ of our $Q_1=Q_2$ solution, with parameter $q$, then describes the 
further twisted compactification of this four-dimensional theory on the spindle $\Sigma$, with $0<q<\frac{1}{2}(n_--n_+)$ 
determining the twisting. The central charge of these  gravity solutions is
\begin{align}\label{cM5intro}
c \, = \, \frac{4q^2(n_- - n_+ - 2 q)}{ n_- n_+ [n_- (n_+ + 2 q) - q (2 n_+ + 3 q)]} (g-1) N^3 \, .
\end{align}
We again reproduce this result from a dual field theory calculation, 
this time utilizing the anomaly polynomial of the 
theory on $N$ M5-branes, doubly wrapped on $\Sigma_g\times \Sigma$.

The outline of the paper is as follows. In section \ref{sec:sugra} 
we construct the AdS$_3\times Y_7$ solutions of interest by 
first constructing $d=5$ AdS$_3\times \Sigma$ solutions 
in $U(1)^3$ gauged supergravity, and then imposing 
appropriate quantization conditions on the gauge field fluxes 
through the spindle $\Sigma=\mathbb{WCP}^1_{[n_-,n_+]}$ so that the uplift on $S^5$ leads to a smooth 
seven-manifold $Y_7$.  We also compute the central charge and R-symmetry gauge 
field for these gravity solutions. In section \ref{sec:M5} 
we similarly construct AdS$_3\times Y_8$ solutions 
to M-theory by uplifting the same $d=5$ solutions, but now necessarily with $Q_1=Q_2$,  
and compute the central charge.
In section \ref{sec:fieldtheory} we 
compute the $d=2$ anomaly polynomial for the twisted compactification 
of $\mathcal{N}=4$ SYM on $\Sigma$, and using $c$-extremization then 
compute the exact central charge and superconformal R-symmetry in field theory, 
finding agreement with the gravity duals. We perform 
a similar computation for the  theory on $N$ M5-branes wrapped
on $\Sigma_g\times \Sigma$, again finding agreement with the 
central charge computed from the supergravity solution. 
We conclude in section \ref{sec:discuss} with a brief discussion.

\

\noindent {\bf Note added}: As this paper was being completed we learned of the work \cite{Hosseini:2021fge}, which  has overlap with the D3-brane results presented here.

\section{D3-brane supergravity solutions}\label{sec:sugra}

In this section we construct a family of supersymmetric 
AdS$_3$ solutions to $d=5$, $U(1)^3$ gauged supergravity. 
These uplift on an $S^5$ internal space to corresponding 
AdS$_3\times Y_7$ solutions of type IIB supergravity. 
The local form of the solutions was first presented in
\cite{Gauntlett:2006ns}, where they were obtained from 
multi-charge superstar solutions in \cite{Cvetic:1999xp}. Shortly afterwards the same solutions 
were rediscovered in \cite{Kunduri:2007qy}, where they 
arise as the uplifts of near-horizon limits of a family $d=5$ 
unbalanced black ring solutions. The novelty 
of this section is the global analysis that we perform for the allowed 
values of the parameters in the solution, 
which is very different to the  approach taken in \cite{Kunduri:2007qy} and
\cite{Gauntlett:2006ns}. 
Instead we
follow \cite{Ferrero:2020laf}. 
The resulting description will be 
crucial to 
match the dual field theory computation in section \ref{sec:fieldtheory}.

\subsection{Local form of the solutions}\label{sec:local}

The action of $d=5$, $U(1)^3$ gauged supergravity is
\be
\begin{split}
S & \, = \, \frac{1}{16\pi G_{(5)}}\int \Big\{ \diff^5 x \sqrt{-\det g}\,  \Big[
R + 4\sum_{I=1}^3 (X^{(I)})^{-1} - \frac{1}{2}\sum_{I=1}^3 (X^{(I)})^{-2}(\partial X^{(I)})^2 \\
& \qquad \qquad \qquad \qquad \qquad \qquad   - \frac{1}{4}\sum_{I=1}^3(X^{(I)})^{-2}(F^{(I)})^2\Big]
- F^{(1)}\wedge F^{(2)} \wedge A^{(3)}\,\Big\} .
\end{split}
\ee
Here $A^{(I)}$, $I=1,2,3$, are the three $U(1)$ gauge fields, with field strengths $F^{(I)}=\diff A^{(I)}$, 
and the three scalar fields $X^{(I)}$ satisfy the constraint $X^{(1)}X^{(2)}X^{(3)}=1$. 
We have set the gauge coupling constant to $1$. A consistent truncation 
to minimal gauged supergravity may be obtained by 
setting the scalars $X^{(I)}=1$ and the three gauge fields 
equal $A^{(I)}=A$, $I=1,2,3$, where  to compare to the conventions in 
\cite{Ferrero:2020laf} one should also rescale the gauge field 
$A$ by factor of $\frac{3}{2}$.

Our starting point is the following supersymmetric AdS$_3$ solution to this theory \cite{Kunduri:2007qy}:
\be
\begin{split} \label{5dmetric}
\diff s^2_5 & \,  = \,  H(x)^{1/3} \left[\diff s^2_{\mathrm{AdS}_3} + \frac{1}{4P(x)} \diff x^2 +\frac{P(x)}{H(x)}\diff\phi^2\right]\, ,\\
A^{(I)}&\, = \, \frac{x-\alpha}{x+3K_I} \diff \phi\, ,\\
 X^{(I)} &\, =\,  \frac{H(x)^{1/3}}{x+3K_I}\, .
\end{split}
\ee
Here $\diff s^2_{\mathrm{AdS}_3}$ a unit radius metric on AdS$_3$, the metric functions are the polynomials
\be\label{metricfunctions}
\begin{split}
H(x) & \, \equiv \,  (x + 3K_1)(x + 3K_2)(x + 3K_3) \, = \,  x^3 + 3 c_1 x + c_2\, ,\\
	P(x) &\, \equiv \,  H(x) - (x-\alpha)^2\, , \\
\end{split}
\ee
and $\alpha$, $K_I$ are constants, with the latter satisfying the constraint
$K_1+K_2+K_3=0$. In the first line of \eqref{metricfunctions} 
we have  introduced the elementary symmetric polynomials 
for the parameters $K_I$, namely  $c_1 \equiv 3(K_1 K_2 + K_2 K_3 + K_1 K_3)$, $c_2 \equiv 27 K_1 K_2 K_3$. 
We note that the solution to minimal gauged supergravity studied 
in \cite{Ferrero:2020laf} is obtained by setting $K_I=0$ for all $I=1,2,3$, 
which is then parametrized by the single parameter~$\alpha$.

Any solution of $d=5$, $U(1)^3$ gauged supergravity may be uplifted (locally)
on an $S^5$ internal space to a solution of type IIB supergravity 
\cite{Cvetic:1999xp}. The ten-dimensional metric~is
\begin{align}\label{lift}
L^{-2}\diff s^2_{10}& \, = \,  W^{1/2}\diff s^2_5 +W^{-1/2}\sum_{I=1}^3 (X^I)^{-1} \left[ \diff \mu_I^2 + \mu_I^2 (\diff\phi_I + A^{(I)})^2 \right]\, ,
\end{align}
while the self-dual five-form flux is
\begin{align} \label{fiveform}
L^{-4} F_5  \, = \,  (1 + \star_{10}) \Big\{ &2 \sum_{I=1}^3 [(X^{(I)})^2 \mu_I^2 - W X^{(I)}] \vol_5 + \frac{1}{2} \sum_{I=1}^3 (X^{(I)})^{-1} \star_5 \diff X^{(I)} \wedge \diff (\mu_I^2)  \nonumber \\
 & + \frac{1}{2 } \sum_{I=1}^3 (X^{(I)})^{-2} \diff (\mu_I^2) \wedge (\diff \phi_I + A^{(I)}) \wedge \star_5 F^{(I)} \Big\}\, .
\end{align}
Here we have introduced an arbitrary length scale parameter $L>0$ into the form of the 
solution, which will be fixed via flux quantization later.  We have denoted 
the volume form on $\diff s^2_5$ by $\vol_5$, and
 $\star$ denotes
the Hodge duality operator in the appropriate dimension. As in \cite{Gauntlett:2006ns}
we have corrected the sign of the second factor in \eqref{fiveform}, 
compared with the expression given
 in \cite{Cvetic:1999xp}. We have  also introduced the warp factor function
\begin{align}\label{warp}
W\, = \, \sum_{I=1}^3 \mu_I^2 X^{(I)} \, > \, 0 \, .
\end{align}
Finally, $\{\mu_I, \phi_I\}$ form a system of polar coordinates on $S^5\subset \R^2\oplus\R^2\oplus\R^2$, 
where correspondingly $\sum_{I=1}^3\mu_I^2=1$ and the angular coordinates $\phi_I$ 
each have period $2\pi$. Notice that the warp factor function \eqref{warp} 
depends on both the internal $S^5$ coordinates $\mu_I$ and {\it a priori} on the $d=5$ coordinates, via the 
scalar fields $X^{(I)}$.

Applying the uplifting formula \eqref{lift} to the AdS$_3$ solution \eqref{5dmetric} gives rise to 
the following warped AdS$_3\times Y_7$ metric
\begin{align}\label{3+7}
\diff s^2_{10}& \, = \, L^2W^{1/2}H(x)^{1/3} \left(\diff s^2_{\mathrm{AdS}_3} + \diff s^2_{Y_7}\right)\, ,
\end{align}
where the seven-dimensional metric on $Y_7$ is
\be
\begin{split}\label{7dmetric}
\diff s^2_{Y_7} &  \, = \, \frac{\diff x^2}{4P(x)} +   \frac{P(x)}{H(x)} \diff\phi^2\\
& \quad + \frac{1}{W H(x)^{2/3}}\sum_{I=1}^3 (x + 3K_I) \left[\diff \mu_I^2+\mu_I^2 \left(\diff \phi_I + \frac{(x-\alpha)}{x+3K_I}\diff \phi \right)^2 \right]\, .
\end{split}
\ee
One can verify that this is the same metric as that given in section 5.2 of \cite{Gauntlett:2006ns}, 
where we identify their coordinates and parameters, in terms of those presented here, as $w=x-\alpha$, $z=-2\phi$, $q_I=\alpha+3K_I$. 
In particular, reference \cite{Gauntlett:2006ns} also shows that the solution is supersymmetric, 
with the dual $d=2$ CFT having $(0,2)$ supersymmetry. 

\subsection{Global analysis and general solution}\label{sec:global}

Given a local AdS$_3\times Y_7$  solution of the form \eqref{7dmetric}, one would 
like to choose the parameters $\alpha, K_I$ and ranges of coordinates so as to obtain 
a smooth metric on a compact internal space $Y_7$. An analysis of this 
was carried out in  \cite{Gauntlett:2006ns} (see also \cite{Kunduri:2007qy}), 
where one looks at the degeneration loci of Killing vector fields on $Y_7$, 
which are constant linear combinations of the four vector fields $\partial_\phi$ and $\partial_{\phi_I}$, $I=1,2,3$. 
We instead follow the approach of \cite{Ferrero:2020laf}, 
first obtaining a suitably regular $d=5$ 
AdS$_3\times \Sigma$ solution, where $\Sigma=\mathbb{WCP}^1_{[n_-,n_+]}$ is a two-dimensional 
orbifold known as a spindle. We then appropriately quantize 
the $U(1)^3$ gauge field fluxes so as to obtain a regular fibration of $S^5$ 
over this five-dimensional solution. 

We begin then with the $d=5$ metric in \eqref{5dmetric}, which we rewrite as 
\begin{align}
\diff s^2_5 & \,  = \,  H(x)^{1/3} \left(\diff s^2_{\mathrm{AdS}_3} + \diff s^2_\Sigma\right)\, ,
\end{align}
with the two-dimensional surface $\Sigma$ having coordinates $x$, $\phi$.
In order for the metric to have the correct signature and be non-singular we 
would like to choose 
the parameters so that $H(x)>0$ is strictly positive, and similarly
$P(x)\geq 0$. In order that $\Sigma$ forms a compact surface without 
boundary, we furthermore assume that 
$x\in [x_1,x_2]$ lies between two roots 
of the polynomial $P(x)$ in \eqref{metricfunctions}.  
In a certain regime of parameters, that we make explicit 
below, we find that the three roots $x_i$, $i=1,2,3$, of $P(x)$ are real 
and positive, and we then take $0<x_1<x_2<x_3$ so that $P(x)\geq 0$ 
for $x\in [x_1,x_2]$. 
Defining $\varrho_i = 2 (x-x_i)^{1/2}$ for $i=1,2$ we find that as $x \rightarrow x_i$  the metric on the two-dimensional 
surface $\Sigma$ approaches
\begin{align}
\diff s^2_{\Sigma}\,  \simeq\,  \frac{1}{4  P'(x_i)} \left(\diff\varrho_i^2+\CC_i^2 \varrho_i^2 \, \diff \phi^2\right)\,, \qquad \mbox{where}\quad   \CC_i \, \equiv \,   \left|\frac{P'(x_i)}{x_i - \alpha}\right|\,, \quad  i\, = \, 1,2\, .
\end{align}
As for the solution in \cite{Ferrero:2020laf}, it is not possible to remove both conical singularities 
at the roots $x=x_i$ by making a single choice $\Delta\phi$ of the period for $\phi$. Instead we 
impose
\begin{align}\label{delwcond}
\Delta \phi\, = \, \frac{2\pi}{\CC_1 n_+}\, = \, \frac{2\pi}{\CC_2 n_-}\, ,
\end{align}
where $n_\pm\in \mathbb{N}$. The resulting surface $\Sigma=\mathbb{WCP}^1_{[n_-,n_+]}$
is then an orbifold known as a spindle. This is topologically a two-sphere, but with 
conical deficit angles 
$2\pi(1-\frac{1}{n_\pm})$ at the poles $x=x_1$, $x_2$. 

After a little computation we find that the second equality in \eqref{delwcond} may be written as
\begin{equation} \label{quotweights2}
\frac{n_-}{n_+} \, = \,  \left| \frac{x_2-\alpha}{x_1-\alpha} \right| \left| \frac{-1+2x_1+x_2}{-1+x_1+2x_2} \right| \, ,
\end{equation}
where the roots $x_1, x_2$ satisfy the equations 
\be
\begin{split} \label{ceq}
2\alpha + 3c_1 & \, = \, x_1+x_2 -x_1^2-x_2^2-x_1 x_2\,,\\
c_2 & \, = \, \alpha^2+x_1^2x_2+x_1x_2^2-x_1x_2\, ,
\end{split}
\ee
and we have eliminated $x_3$ via the relation $x_1+x_2+x_3=1$. 

We will also need the fluxes of the gauge field strengths $F^{(I)}=\diff A^{(I)}$ through the surface $\Sigma$. 
Using the expression for $A^{(I)}$ in \eqref{5dmetric} we compute
\begin{equation}
	\label{fluxes}
	\begin{split}
		Q_I &\, \equiv \,  \frac{1}{2\pi} \int_\Sigma F^{(I)} \, = \,  \frac{(x_2 - x_1) (\alpha + 3K_I)}{(x_1 + 3K_I) (x_2 + 3K_I)}\frac{\Delta\phi}{2\pi} \,  .
	\end{split}
\end{equation}
As explained in \cite{Ferrero:2020laf}, \cite{Ferrero:2020twa}, for $\Sigma=\mathbb{WCP}^1_{[n_-,n_+]}$ 
the appropriate quantization condition on these fluxes is that 
\begin{align}\label{quantize}
Q_I \, \equiv \, \frac{p_I}{n_-n_+}\, , \qquad \mbox{where}\quad p_I\in  \Z\, .
\end{align}
Specifically, the circles parametrized by $\phi_I$ inside the internal space $S^5\subset \C^3$ then  give
a well-defined orbifold circle fibration over $\Sigma$. Moreover, provided the integers $p_I$ are coprime to 
both of $n_\pm$, the total space of this fibration is smooth.\footnote{See appendix A of 
\cite{Ferrero:2020twa} for a detailed discussion of this.} This leads to a 
fibration
\begin{align}\label{fibre}
S^5\ \hookrightarrow \ Y_7 \ \rightarrow  \ \Sigma \, = \, \mathbb{WCP}_{[n_-,n_+]}\, ,
\end{align}
where the total space $Y_7$ is a compact seven-manifold, and  the twisting 
is determined by the integers $p_I$. 

At this stage we have a three-parameter family of solutions, determined by 
the constants $\alpha, K_1, K_2$, where the roots $x_1,x_2$ obey 
\eqref{ceq}. We would like to express the solution in terms 
of the more physical flux parameters $p_I$ and $n_\pm$, the ratio of the latter determined 
via \eqref{quotweights2}. It turns out to be convenient to express
quantities in terms of two of the three fluxes, which we take to be 
\begin{align}
p_1 \, \equiv \, q\, , \qquad p_2 \, \equiv \, p\, .
\end{align}
After some work, it is possible to solve the three equations \eqref{quotweights2}, and \eqref{ceq} for $\alpha$, $K_1$, $K_2$, $x_1,$ and $x_2$  in terms of $n_\pm$, together with $q$ and $p$. We find the rather unwieldy expressions
\begin{align}\label{soln}
x_1 &\, = \, \frac{1}{\mathcal{D}} \Big\{(p^2+p q+q^2)^2+3 n_+ p q (p+q+n_+)+2 (n_+-n_-) (p+q) (p^2+q^2) \nn
& \qquad +(n_-^2+n_+^2) (p^2+q^2)-n_- n_+ (3 p^2+4 p q+3 q^2)+n_- n_+ (n_--n_+) (p+q)\Big\} \:, \nn[10pt]
x_2 &\, = \, \frac{1}{\mathcal{D}} \Big\{(p^2+p q+q^2)^2-3 n_- p q (p+q-n_-)+2 (n_+-n_-) (p+q) (p^2+q^2) \nn
& \qquad +(n_-^2+n_+^2) (p^2+q^2)-n_- n_+ (3 p^2+4 p q+3 q^2)+n_- n_+ (n_--n_+) (p+q)\Big\} \:,\nn[2pt]
&\\[-7pt]
K_1 & \, =\,  \frac{1}{3\mathcal{D}} \Big\{3 p^2 (p+q)^2-(p^2+p q+q^2)^2+2 (n_--n_+) (q^2-2 p^2) (p+q)-2n_- n_+ p q\nn
&\qquad\qquad -(n_-^2+n_+^2) (q^2-2 p^2)+3n_- n_+ (q^2-2p^2)-n_- n_+ (n_--n_+) (q-2 p)\Big\} \:,\nn[10pt]
K_2 &\, =\,  \frac{1}{3\mathcal{D}} \Big\{3 q^2 (p+q)^2-(p^2+p q+q^2)^2+2 (n_--n_+) (p^2-2 q^2) (p+q)-2n_- n_+ p q\nn
&\qquad\qquad -(n_-^2+n_+^2) (p^2-2 q^2)+3n_- n_+ (p^2-2q^2)-n_- n_+ (n_--n_+) (p-2 q)\Big\} \:,\nonumber
\end{align}
where we have defined the denominator term
\begin{align}
\mathcal{D}\, \equiv\,  3\, [(n_- - n_+)(p+q)+n_-n_+-p^2-p q-q^2]^2\, .
\end{align}
The parameter $\alpha$ may then be determined from the first equation 
in \eqref{ceq}. 

The third flux is computed to be 
\begin{equation}
	Q_3\, = \, \frac{n_- - n_+ - q - p}{n_- n_+}\, ,
\end{equation}
which notice also satisfies the quantization condition \eqref{quantize}, with 
\begin{align}
p_1 \, = \, q\, , \qquad p_2\, = \, p\, , \qquad p_3\, = \, n_--n_+-q-p\, .
\end{align}
We then also have
\begin{align}\label{chargesum}
Q_1 + Q_2 + Q_3 \, = \, \frac{n_--n_+}{n_-n_+}\, .
\end{align}
We find that the roots obey the assumed inequalities $0 < x_1 < x_2 < x_3$ provided 
\begin{align}\label{inequalities}
q, \, p \, > \, 0\, , \qquad q+p\, < \, n_--n_+\, .
\end{align}

We now have a family of regular supersymmetric AdS$_3\times Y_7$ solutions, with $Y_7$ having the fibration
structure \eqref{fibre}, parametrized by the integers $n_-,n_+,p$ and $q$. 
We address quantization of the five-form flux $F_5$ in the next subsection.

\subsection{Central charge and R-symmetry}\label{sec:central}

The AdS$_3$ solutions are, via the AdS/CFT correspondence, expected to be dual 
to $d=2$, $(0,2)$ SCFTs. The central charge of these CFTs is computed in gravity using 
the standard formula \cite{Brown:1986nw}
\be
\label{ccharge}
	c \, = \,  \frac{3L}{2G_{(3)}}\, ,
\ee
where $G_{(3)}$ is the effective Newton constant in three dimensions. In turn the 
latter is computed via 
 dimensional 
reduction on $Y_7$, and  is given by \cite{Couzens:2018wnk}
\begin{equation}
	\frac{1}{G_{(3)}} \, = \,  \frac{L^7}{G_{(10)}} \int_{Y_7} W^2 H(x)^{4/3} \, \text{vol}_7 \, = \,  \frac{\pi^3 L^7 \Delta\phi}{2 G_{(10)}} \, (x_2-x_1) \: ,
\end{equation}
where we have explicitly evaluated the integrals over $S^5$ and the spindle $\Sigma$.
The ten-dimensional Newton constant is
\begin{equation}
	G_{(10)} \, = \,  \frac{(2\pi)^7 g_s^2 \, \ell_s^8}{16\pi} \:, 
\end{equation}
where $\ell_s$ and $g_s$ are the constant string length and string coupling constant, respectively. 
The central charge \eqref{ccharge} is then
\be
\label{cchargeagain}
	c \, =  \, \frac{3 L^8 \Delta\phi}{32\pi^3 g_s^2 \, \ell_s^8} \, (x_2-x_1) \:.
\ee

In order to obtain a good string theory solution we must also quantize the 
flux of the closed five-form \eqref{fiveform} through five-cycles in 
the ten-dimensional spacetime. In particular, integrating $F_5$ through a copy 
of the $S^5$ fibre, at any point in the $d=5$ spacetime, we find  the total flux is
\begin{equation}
N \, \equiv \, \frac{1}{(2\pi \ell_s)^4 g_s} \int_{S^5} F_5  \, =  \, \frac{L^4}{4\pi \ell_s^4 \, g_s}\, .
\end{equation}
Eliminating $L$ in the central charge \eqref{cchargeagain} using this last equation then gives
\begin{equation}\label{cchargeneat}
c \, = \,  \frac{3 }{2 \pi} \Delta \phi (x_2-x_1)N^2\,.
\end{equation}
Finally, substituting the solution \eqref{soln} into \eqref{cchargeneat}, we find the remarkably simple expression
\begin{equation} \label{cgrav3charge}
	c \, =  \, \frac{3 p q (n_- - n_+ - q - p)}{n_- n_+ [(n_--n_+) (p+q) + n_-n_+-p^2-p q-q^2]} N^2 \,.
\end{equation}
This is our final formula for the central charge, parametrized in terms of the integers 
$n_-, n_+,  p,q$ and $N$. It will be useful to note that \eqref{delwcond} gives the 
period $\Delta\phi$ of $\phi$ to be 
\begin{align}\label{Deltaphi}
\frac{\Delta\phi}{2\pi}\, = \, \frac{ (n_- - n_+)(p+q)+n_-n_+-p^2-p q-q^2}{ n_-n_+( n_-+ n_+)}\, ,
\end{align}
an expression we will need momentarily.

The $U(1)$ R-symmetry of the dual $d=2$, $(0,2)$ field theory is realized in the gravity 
solution as a Killing vector field $R_{\mathrm{2d}}$ of the AdS$_3\times Y_7$ 
gravity solution. This is defined as a certain bilinear in the Killing spinors 
on $Y_7$ \cite{Gauntlett:2007ts}, and combining the general results 
of the latter reference with the form of the solution given in \cite{Gauntlett:2006ns} 
we can read off that this vector field is simply
\begin{align}\label{Rvec}
R_{\mathrm{2d}}\, = \, \partial_\phi\, .
\end{align}
Here we have normalized the R-symmetry so that the Killing spinor on $Y_7$ has unit 
charge under the Lie derivative along this vector field, giving an explicit 
phase dependence of $\ex^{\ii \phi}$. 
On the other hand, this Killing spinor arises as a tensor product 
of the $d=5$ Killing spinor with a spinor on the internal $S^5$. 
This ansatz preserves an $\mathcal{N}=1$ spinor, out of the full 
$\mathcal{N}=4$ supersymmetry of $S^5$, with the spinor on $S^5$ having
charge $\frac{1}{2}$ under each of the three vector fields $\partial_{\phi_I}$
generating the $U(1)^3$ isometry of $S^5\subset \C^3$. 
As discussed in \cite{Ferrero:2020laf}, we would like to 
choose a gauge for the $d=5$ gauge fields such that the
Killing spinor is uncharged under the $U(1)$ isometry 
that rotates the spindle.
Since  a gauge transformation of the $A^{(I)}$ precisely 
leads to a phase rotation in the corresponding Killing spinor, which has 
unit charge under each $A^{(I)}$, this may be achieved via the  gauge/coordinate transformation
\begin{align}\label{coordtrans}
	\tilde{\phi}_I \, =\,  \phi_I + \tfrac{2}{3} \,  \phi\, , \qquad \tilde{A}^{(I)} \, = \, A^{(I)} - \tfrac{2}{3} \, \diff \phi\, ,
\end{align}
which ensures that $\diff\phi_I+A^{(I)}=\diff\tilde{\phi}_I+\tilde{A}^{(I)}$. 
Defining also
\begin{align}
\varphi \, \equiv \,  \frac{2\pi}{\Delta\phi} \, \phi\, ,
\end{align}
so that $\Delta\varphi = 2\pi$, and completing the coordinate transformation \eqref{coordtrans} by defining
\begin{align}\label{tildephi}
\tilde{\varphi}\, = \ \varphi\, ,
\end{align}
we see that the R-symmetry vector \eqref{Rvec} is 
\begin{equation}\label{premixing}
	\begin{split}
		R_{\mathrm{2d}} &\, = \,  \partial_\phi \, =  \,  \frac{2\pi}{\Delta\phi} \, \partial_\varphi \, = \,   \frac{2\pi}{\Delta\phi} \left(\sum_{I=1}^3 \frac{\partial \tilde{\phi}_I}{\partial \varphi} \, \partial_{\tilde{\phi}_I} + \frac{\partial \tilde{\varphi}}{\partial \varphi} \, \partial_{\tilde{\varphi}}\right) \, = \,  \frac{2}{3} \sum_{I=1}^3  \partial_{\tilde{\phi}_I} + \frac{2\pi}{\Delta\phi} \, \partial_{\tilde{\varphi}}\, .
\end{split}
\end{equation}
Note that $\partial_{\tilde{\varphi}}$ generates the $U(1)$
isometry of the spindle, that we refer to as $\UJ$ in the subsequent discussion. 
The gauge transformation \eqref{coordtrans} leads to a phase $\ex^{\ii (\tilde{\phi}_1+\tilde{\phi}_2+\tilde{\phi}_3)/2}$
in the Killing spinor on $Y_7$, but this is now invariant under $\UJ$.  We may then identify the Killing vector field on $S^5$
\begin{equation}
R_{\mathrm{4d}}\, = \, \frac{2}{3} \sum_{I=1}^3  \partial_{\tilde{\phi}_I}\, ,
\end{equation}
with the superconformal $\mathcal{N}=1$ R-symmetry before compactification 
of the theory on $\Sigma$, and from \eqref{premixing} hence write
\begin{equation}\label{mixing}
R_{\mathrm{2d}} \, = \, R_{\mathrm{4d}} + \frac{n_-n_+( n_-+ n_+)}{ (n_- - n_+)(p+q)+n_-n_+-p^2-p q-q^2}\, \partial_{\tilde{\varphi}}\, .
\end{equation} 
Here in the second term we have substituted for $\Delta\phi$ using \eqref{Deltaphi}. 
Equation \eqref{mixing} states that the $d=4$ $U(1)$ R-symmetry 
mixes with $\UJ$ in flowing to the $d=2$ 
$U(1)$ R-symmetry in the IR. 
 We shall recover 
this formula from a dual field theory calculation 
in section~\ref{sec:fieldtheory}, along with the central charge 
\eqref{cgrav3charge}. 

\subsection{Special cases}\label{sec:cases}

In this section we briefly analyse some interesting special cases of the general 
solutions of section \ref{sec:global}, in particular making contact with \cite{Ferrero:2020laf}. 

Setting $p=q$ implies that the fluxes $Q_1=Q_2$ are equal, and 
hence also $A^{(1)}=A^{(2)}$, and $X^{(1)}=X^{(2)}$. In this case 
\eqref{soln} simplifies~to
\be\label{twosoln}
\begin{split}
x_1 &\, = \,  \frac{q\,(n_+ + q) [2 n_-^2 - 2 n_- (n_+ + 4 q) + q\, (5 n_+ + 9 q)]}{3\, [n_- (n_+ + 2 q) - q\,(2 n_+ + 3 q) ]^2} \,,\\[6pt]
x_2 &\, = \,  -\frac{q\,(n_- - q)[2 n_+^2 - 2 n_+ (n_- - 4 q) - q\, (5 n_- - 9 q)]}{3\, [n_- (n_+ + 2 q) - q\,(2 n_+ + 3 q) ]^2} \,,\\[6pt]
K_1 &\, = \, K_2\, = \,  \frac{q\,(n_- - n_+ - 3 q) (n_+ + q) (n_- - q)}{9\, [n_- (n_+ + 2 q) - q\,(2 n_+ + 3 q) ]^2} \,,
\end{split}
\ee
with corresponding central charge
\begin{equation} \label{cgrav}
c \, = \, \frac{3 q^2 (n_- - n_+ - 2q)}{n_- n_+ [n_- (n_+ + 2 q) - q\,(2 n_+ + 3 q) ]}N^2  \,.
\end{equation}
This sub-family of solutions also uplift to solutions of $d=11$ supergravity, 
as discussed in section \ref{sec:M5}. 

Finally, setting all three charges equal gives, from \eqref{chargesum},
\begin{align}\label{equals}
p \, = \, q\, = \, \frac{1}{3}(n_--n_+)\, .
\end{align}
This is precisely the solution to minimal gauged supergravity studied in \cite{Ferrero:2020laf}, 
where we note that $p,q\in \Z$ then requires $n_--n_+$ to be divisible by 3, as discussed in \cite{Ferrero:2020laf}.\footnote{In 
the notation of that reference, we take the K\"ahler-Einstein four-manifold to be $\mathrm{KE}_4=\mathbb{CP}^2$, and 
$k=1$ then gives $S^5$ as the internal space. But this then requires $n_--n_+$ to be divisible by the Fano
index of $\mathbb{CP}^2$, which is $I=3$. The parameters $p$ and $q$ in this paper should not be confused 
with those in \cite{Ferrero:2020laf}, and also recall that one should rescale our gauge field 
by a factor of $\frac{3}{2}$ to match to the conventions in \cite{Ferrero:2020laf}.}  
Imposing \eqref{equals}, our central charge \eqref{cgrav3charge} and R-symmetry gauge field 
\eqref{mixing} reduce to the expressions given in \cite{Ferrero:2020laf}.  

\section{M5-brane supergravity solutions}\label{sec:M5} 

In this section we construct a family of supersymmetric AdS$_3\times Y_8$ solutions 
to $d=11$ supergravity. These arise by uplifting the $Q_1=Q_2$ solutions of $d=5$, $U(1)^3$ 
gauged supergravity 
in section \ref{sec:cases} on the Maldacena-N\'u\~nez solution \cite{Maldacena:2000mw}, 
using the consistent truncation of \cite{Gauntlett:2007sm}. We interpret these M-theory solutions
as the near-horizon limit of $N$ M5-branes 
wrapped on $\Sigma_g\times \Sigma$.

\subsection{Romans supergravity and uplift}\label{sec:upliftM5}

There are a number of different consistent truncations of type IIB supergravity on $S^5$, 
among them being both the $U(1)^3$ gauged supergravity of section \ref{sec:local}, 
but also the Romans $\mathcal{N} = 4$, $SU(2) \times U(1)$  supergravity \cite{Romans:1985ps}. 
The latter preserves a different subgroup of the full $SO(6)$ R-symmetry of the internal $S^5$. 
The bosonic sector of the Romans theory contains a scalar field $X$, a triplet of $SU(2)$ gauge fields $B_\mu^i$, $i=1,2,3$, a $U(1)$ gauge field $A_\mu$, and two two-forms. This theory can be further truncated by setting to zero the two-forms and truncating $SU(2)$ to its Abelian subgroup. It was shown in 
\cite{Lu:1999bw} that the resulting theory is the same as the $U(1)^3$ theory with two gauge fields $A^{(1)}=A^{(2)}$ and two scalar fields 
$X^{(1)}=X^{(2)}$ set equal, and in particular then
\begin{equation}\label{trun}
	X \, = \, X^{(1)} \:, \qquad B_\mu^1 \, = \, B_\mu^2 \, = \, 0 \:, \qquad B_\mu^3 \, = \, A^{(1)}_\mu \:, \qquad A_\mu \, = \, A_\mu^{(3)} \:.
\end{equation}

On the other hand, this Romans supergravity theory is also a consistent truncation of $d=11$ supergravity, as shown in \cite{Gauntlett:2007sm}. The vacuum AdS$_5$ solution uplifts to a warped AdS$_5\times N_6$ solution, which for a given
internal space $N_6$ is  dual to an $\mathcal{N}=2$ SCFT in four dimensions. A particular example, studied 
in \cite{Gauntlett:2007sm}, is the Maldacena-N\'u\~nez solution, which is dual to the $d=4$, $\mathcal{N}=2$ SCFT living on M5-branes wrapping a Riemann surface, holomorphically embedded in a Calabi-Yau two-fold. 
More generally there are the AdS$_5\times N_6$ solutions of \cite{Gaiotto:2009gz}, corresponding to M5-branes 
wrapping a Riemann surface with punctures, and 
 holographically dual 
to $d=4$, $\mathcal{N}=2$ SCFTs of class $\mathcal{S}$.  Our five-dimensional 
solutions with $Q_1=Q_2$, given by \eqref{twosoln}, may 
thus also be uplifted to interesting classes of M5-brane solutions.
Specifically, the M5-brane is wrapped on a Riemann surface (in general with punctures), which 
is then further wrapped on the spindle $\Sigma=\mathbb{WCP}^1_{[n_-,n_+]}$ 
to obtain a two-dimensional theory. 

In this section 
for simplicity we focus on the uplift on the Maldacena-N\'u\~nez solution. 
A solution to the $d=5$ Romans theory with scalar field $X$, and Abelian 
gauge fields $B^3=A^{(1)}$, $A=A^{(3)}$, with $B^1=B^2=0$ as in \eqref{trun}, 
uplifts to a $d=11$ solution with metric \cite{Gauntlett:2007sm}\footnote{We note that in \cite{Gauntlett:2007sm} 
the uplifting formula is stated for $d=5$ gauge coupling set to $m_{\mathrm{there}}=\frac{1}{2}$, while our $d=5$ solutions 
have this quantity set to 1. We thus need to rescale our fields, and in the notation 
of \cite{Gauntlett:2007sm} correspondingly $A^3_{\mathrm{there}}=2\sqrt{2}
A^{(1)}$, $B_{\mathrm{there}}=2A^{(3)}$.}
  \begin{equation}\label{11d}
	\begin{split}
		L^{-2} \diff s_{11}^2 &= 2^{-2/3} \Omega^{1/3} \diff s_5^2 +2^{1/3} X \Omega^{1/3} \left(\diff \theta^2 + \diff s_{\Sigma_g}^2\right) \\
		& + 2^{1/3} X \Omega^{-2/3} \sin^2\theta \left(\diff \psi + V +  A^{(3)}\right)^2 + \frac{2^{-2/3}}{X^2} \, \Omega^{-2/3} \cos^2\theta D\mu_i D\mu_i \:. 
	\end{split}
\end{equation}
Here $\diff s^2_5$ denotes the five-dimensional gauged supergravity metric, $\diff s^2_{\Sigma_g}$ is the metric on 
 a unit radius hyperbolic plane, quotiented by a discrete group of isometries 
to obtain a compact Riemann surface of genus $g>1$, and  we have introduced the warp factor function
\begin{align}
\Omega\, \equiv  \, \cos^2\theta + \frac{1}{2 X^3} \,\sin^2\theta\, .
\end{align}

The coordinates 
 $\theta$, $\psi$, and $\mu_i$, $i=1,2,3$, are (constrained) coordinates on $S^4 \subset \mathbb{R}^2 \oplus \mathbb{R}^3$, 
where $\theta\in [0,\frac{\pi}{2}]$ describes the polar direction in the latter splitting. 
Smoothness of the metric \eqref{11d} at $\theta=0$ fixes the period $\Delta\psi=2\pi$, 
and the local one-form $V$ on $\Sigma_g$ is such that $\diff V = - \vol_{\Sigma_g}$. We note that
\begin{align}\label{VolSigmag}
\Vol(\Sigma_g)\, = \, \int_{\Sigma_g} \vol_{\Sigma_g}\, = \, 4\pi (g-1) \, = \ -2\pi\,  \chi(\Sigma_g)\, ,
\end{align}
with $\chi(\Sigma_g)$ the Euler number of $\Sigma_g$. This identifies 
the $\R^2$ bundle over $\Sigma_g$, with unit circle in $\R^2=\C$ having 
coordinate $\psi$, as $T^*\Sigma_g$.  This is a local 
Calabi-Yau two-fold, with the $d=11$ solution describing the near-horizon limit of a stack of
M5-branes wrapped on the zero-section $\Sigma_g$.
The M5-branes have corresponding normal space
$\R^5=\C\oplus\R^3$, and denoting the first factor by the 
complex  line bundle 
 $\mathcal{N}_1$, we see that
\begin{align}
\mathcal{N}_1 \, = \, T^*\Sigma_g\otimes L_3\, , 
\end{align}
where $A^{(3)}$ is a connection on a complex line bundle $L_3$ over the $d=5$ spacetime. 
For our solution with $Q_1=Q_2$ in \eqref{twosoln}, this 
is a line bundle over the spindle $\Sigma=\mathbb{WCP}^1_{[n_-,n_+]}$ 
with charge $Q_3$.

Finally, $\mu_i$, $i=1,2,3$,  are constrained coordinates describing a round $S^2\subset \R^3$, 
with $\sum_{i=1}^3\mu_i^2 = 1$, and the connection terms are given by
\begin{align}
	D\mu_1 &\, = \,  \diff \mu_1 +2  A^{(1)} \mu_2 \:,\qquad 
	D\mu_2 \, =\,  \diff \mu_2 -2  A^{(1)} \mu_1 \:,  \qquad 
	D\mu_3 \, = \,  \diff \mu_3 \:.
\end{align}
Writing $\mu_1=\sin\vartheta\cos\nu$, $\mu_2=\sin\vartheta\sin\nu$, $\mu_3=\cos\vartheta$ in spherical polar coordinates, 
the twisted metric on $S^2$ is 
\begin{align}
D\mu_i D\mu_i \, = \, \diff\vartheta^2 + \sin^2\vartheta(\diff\nu - 2A^{(1)})^2\, . 
\end{align}
Since $\nu$ has $\Delta\nu=2\pi$, we see that writing
$\R^3=\C\oplus\R$, the complex line $\C$ is twisted via $L_1^2$, with connection $2A^{(1)}$, so we have
\begin{align}
\mathcal{N}_2 \, = \, L_1^2\, .
\end{align}
The total normal bundle of the M5-branes is then $\mathcal{N}_1\oplus\mathcal{N}_2\oplus \R$, with 
$\R$ the $z$-axis direction in $\R^3$. 

The M-theory four-form $G_4$ may also be read off from the expression in 
\cite{Gauntlett:2007sm}, although we won't need its explicit form in what follows. 

\subsection{Central charge}\label{sec:M5central}

We begin by writing 
the metric \eqref{11d} as
\begin{equation}
	\diff s_{11}^2 \, = \,  \left(16\,\Omega H(x)\right)^{1/3} \left(\diff s^2_{\mathrm{AdS}_3} + \diff s^2_{Y_8}\right) \:,
\end{equation}
where we have uplifted the $d=5$ solution of section \ref{sec:cases} with $Q_1=Q_2$. 
The  eight-dimensional metric on $Y_8$ is then
\begin{equation}
	\begin{split}
		\diff s^2_{Y_8} &= \frac{1}{4P(x)} \, \diff x^2 + \frac{P(x)}{H(x)} \, \diff\phi^2 + \frac{ X}{2H(x)^{1/3}}\left(\diff \theta^2 + \diff s_{\Sigma_g}^2\right) \\
		&\quad + \frac{ X}{2\Omega H(x)^{1/3}} \sin^2\theta \left(\diff \psi + V + A^{(3)}\right)^2 + \frac{1}{4X^2\Omega H(x)^{1/3}} \cos^2\theta D\mu_i D\mu_i \:.
	\end{split}
\end{equation}

The central charge of the dual $d=2$, $(0,\, 2)$ field theories is again given by formula \eqref{ccharge}, where $G_{(3)}$ is computed via dimensional reduction on $Y_8$. We find
\begin{equation}
	\frac{1}{G_{(3)}} \, = \,  \frac{L^8}{G_{(11)}} \int_{Y_8}  \left(16\,\Omega H(x)\right)^{3/2} \vol_8 \, = \,  \frac{16\pi^2 L^8  \Delta\phi}{3 G_{(11)}} \,(x_2-x_1) \Vol(\Sigma_g)\:,
\end{equation}
where in our conventions the eleven-dimensional Newton constant is 
\begin{equation}
	G_{(11)} \, = \,  \frac{(2\pi)^8\, \ell_p^9}{16\pi} \:.
\end{equation}
The central charge is then given by
\begin{equation}\label{cchargeM}
	c \, =\,  \frac{L^9 \Delta\phi}{2\pi^5 \ell_p^9}\,(x_2-x_1)\Vol(\Sigma_g)\:.
\end{equation}

We also need to impose flux quantization  on the four-form in order to have a consistent M-theory solution. Integrating the 
expression for $G_4$ in \cite{Gauntlett:2007sm} over a copy of the $S^4$ fibre we get the number of M5 branes
\begin{equation}
	N \, \equiv \, \frac{1}{(2\pi \ell_p)^3}\int_{S^4} G_4 \, = \,  \frac{L^3}{\pi \ell_p^3} \:.
\end{equation}
Substituting for the length scale $L$ in \eqref{cchargeM}, $\Vol(\Sigma_g)$ given by \eqref{VolSigmag}, 
and also for $x_1$, $x_2$ in \eqref{twosoln} 
and  $\Delta\phi$ in \eqref{Deltaphi} with $p=q$, we find the final central charge
\begin{align}\label{cM5gravity}
c \, = \, \frac{4q^2(n_- - n_+ - 2 q) }{ n_- n_+ [n_- (n_+ + 2 q) - q (2 n_+ + 3 q)]}(g-1) N^3 \, .
\end{align}
This is the central charge of a family of $d=2$, $(0,2)$ theories obtained by 
wrapping $N$ M5-branes on $\Sigma_g\times \Sigma$, with 
$\Sigma=\mathbb{WCP}^1_{[n_-,n_+]}$ a spindle, and the integer $q$ with $0<q<\frac{1}{2}(n_--n_+)$ 
determines the twisting of the normal bundle of the M5-branes. 
We shall recover this from a dual theory calculation in section \ref{sec:fieldtheory}. 

\section{Field theory}\label{sec:fieldtheory}

Although our AdS$_3\times Y_7$ solutions  were first constructed in
\cite{Gauntlett:2006ns}, the description in the latter reference meant that 
the dual $d=2$, $(0,2)$ SCFTs were left unidentified. However, as in 
\cite{Ferrero:2020laf} our alternative $d=5$ construction of the solutions 
leads to a natural conjecture. Specifically, one begins with $\mathcal{N}=4$ 
SYM theory, which is holographically dual to the AdS$_5\times S^5$ 
vacuum of the $d=5$, $U(1)^3$ gauged supergravity theory. 
One then compactifies this theory 
on $\Sigma=\mathbb{WCP}^1_{[n_-,n_+]}$, together with 
background fluxes for the $U(1)^3$ Abelian symmetry 
given by \eqref{quantize}. Recall here 
that $p_1=q$, $p_2=p$, $p_3=n_--n_+-q-p$. 
The solution we have described suggests that this compactification of 
$\mathcal{N}=4$ SYM 
flows to a $d=2$, $(0,2)$ SCFT in the IR. We give 
evidence for this in sections~\ref{sec:anom} and \ref{sec:extreme} by computing the 
central charge and superconformal $U(1)$ R-symmetry 
using purely field theory methods, finding 
precise agreement with the supergravity results.
There is a similar interpretation of the M5-brane solutions in 
section \ref{sec:M5}. In section \ref{sec:anomM5} 
we likewise compactify the theory on $N$ M5-branes 
on $\Sigma_g\times \Sigma$, with appropriate 
background fluxes, and compute the central 
charge of the $d=2$, $(0,2)$ SCFTs, again finding 
agreement with supergravity.

\subsection{D3-brane anomaly polynomial}\label{sec:anom}

Generalizing the approach of \cite{Ferrero:2020laf}, our 
starting point is the anomaly polynomial for 
$\mathcal{N}=4$ SYM, with background gauge field 
fluxes $F^{(I)}$ for the $U(1)^3\subset SO(6)$ Abelian global symmetry group of this theory. 
In the large $N$ limit this anomaly polynomial reads (see, for example, \cite{Hosseini:2020vgl})
\begin{equation} \label{anomaly4d}
\mathcal{A}_{\mathrm{4d}} \, = \, c_1(F^{(1)}) c_1(F^{(2)}) c_1(F^{(3)}) \frac{N^2}{2} \, .
\end{equation}
Here $c_1(F^{(I)})$ denote the first Chern classes of the 
$U(1)$ bundles with  gauge field curvatures $F^{(I)}$, respectively.

We want to compactify this theory on $\Sigma =\mathbb{WCP}^1_{[n_-,n_+]}$, with fluxes given by \eqref{quantize}, and 
compute the anomaly polynomial of the resulting $d=2$ theory.  
For this we also need to take into account the $\UJ$ global symmetry in $d=2$, coming from the $\UJ$ isometry of $\Sigma$. Geometrically, 
this means that we want to compute the anomaly polynomial 
\eqref{anomaly4d}, where the six-manifold $Z_6$ on which it is defined is the total 
space of a $\Sigma$ fibration over 
$Z_4$. We may achieve this by introducing a 
corresponding connection form $\AJ$ for $\UJ$, and 
replacing $\diff \varphi \rightarrow \diff \varphi + \AJ$. 
In doing so it is important that the $d=5$ Killing spinor 
is independent of $\tilde\varphi=\varphi$, which is true in the tilded
gauge defined by \eqref{coordtrans}, \eqref{tildephi}.  
We thus introduce the connection one-forms on $Z_6$:
\begin{equation}\label{AIscript}
\mathscr{A}^{(I)} \,\equiv\,  \left( \frac{x - \alpha}{x + 3K_I} - \frac{2}{3} \right) \frac{\Delta \phi}{2\pi} (\diff \varphi + \AJ) \, \equiv \, \rho_I(x) (\diff \varphi + \AJ)\,, \quad I\, = \, 1,2,3\, ,
\end{equation}
which by construction restricts to the supergravity gauge field $\tilde{A}^{(I)}$ on each $\Sigma$ fibre. We then compute the curvature
\begin{equation}
\mathscr{F}^{(I)}\,  \equiv \, \diff \mathscr{A}^{(I)}\,  = \, \rho_I'(x) \diff x \wedge (\diff \varphi + \AJ) + \rho_I(x) \FJ\,,
\end{equation}
where $\FJ = \diff \AJ$. By construction the integral of $\mathscr{F}^{(I)}$ over a fibre $\Sigma$ gives the flux $Q_I$, as in \eqref{fluxes}. The curvature form $\mathscr{F}^{(I)}$ defines a $U(1)$ bundle $\mathcal{L}_I$ over $Z_6$ by taking $c_1 (\mathcal{L}_I) = [\mathscr{F}^{(I)}/ 2\pi] \in H^2(Z_6,\R)$. We also define $c_1(\JJ) = [\FJ / 2\pi] \in H^2(Z_4,\Z)$. In the anomaly polynomial we then write
\begin{equation}
c_1 (F^{(I)}) \, = \, \Delta_I\,  c_1(R_{\mathrm{2d}}) + c_1 (\mathcal{L}_I)\, ,
\end{equation}
where $R_{\mathrm{2d}}$ is the pull-back of a $U(1)$ bundle over $Z_4$. 
The trial R-charges $\Delta_I$ satisfy the constraint $\Delta_1+\Delta_2+\Delta_3=2$, which 
is equivalent to the superpotential (in $\mathcal{N}=1$ language) having R-charge 2.

The $d=2$ anomaly polynomial is then obtained by integrating $\mathcal{A}_{\mathrm{4d}}$ in \eqref{anomaly4d} over $\Sigma$:
\be
\begin{split} \label{anomalyintegral}
\mathcal{A}_{\mathrm{2d}} & \, =  \, \int_{\Sigma} \mathcal{A}_{\mathrm{4d}} \,  = \, \frac{N^2}{2} \int_{\Sigma} c_1(F^{(1)}) c_1(F^{(2)}) c_1(F^{(3)})\, ,
\end{split}
\ee
which reads
\be\label{A2dfirst}
\begin{split}
\mathcal{A}_{\mathrm{2d}} & \, = \, 
 \frac{N^2}{2} \int_{\Sigma} \Big\{ \Delta_1 \Delta_2 \Delta_3 c_1(R_{\mathrm{2d}})^3 + c_1(R_{\mathrm{2d}})^2 \big[ \Delta_2 \Delta_3 c_1(\mathcal{L}_1) + \Delta_1 \Delta_3 c_1(\mathcal{L}_2) \\
& \qquad \qquad \quad 
+ \Delta_1 \Delta_2 c_1(\mathcal{L}_3) \big] +  c_1(R_{\mathrm{2d}}) \big[\Delta_3 c_1(\mathcal{L}_1) c_1(\mathcal{L}_2) 
+ \Delta_2 c_1(\mathcal{L}_1) c_1(\mathcal{L}_3) \\
& \qquad \qquad  \quad 
 + \Delta_1 c_1(\mathcal{L}_2) c_1(\mathcal{L}_3) \big]+ c_1(\mathcal{L}_1) c_1(\mathcal{L}_2) c_1(\mathcal{L}_3) \Big\}\,. 
\end{split}
\ee
After a computation using the explicit functions $\rho_I(x)$ defined in \eqref{AIscript}, we find
the integral in \eqref{A2dfirst} gives  
\begin{align}\label{anomaly2dfinalfinal}
\mathcal{A}_{\mathrm{2d}}& \, = \,  \frac{N^2}{2} \left\{  \left[ \frac{q \Delta_2 \Delta_3}{n_- n_+} + \frac{p \Delta_1 \Delta_3}{n_- n_+} + \frac{n_- - n_+-p-q}{n_- n_+} \, \Delta_1\Delta_2\right] c_1(R_{\mathrm{2d}})^2 \right. \nn [4pt]
& \qquad \left.+\frac{\Delta_1 f_1(n_-,n_+,p,q)+\Delta_2 f_2(n_-,n_+,p,q)+\Delta_3 f_3(n_-,n_+,p,q)}{3 n_-^2 n_+^2 (n_- + n_+)}\, c_1(R_{\mathrm{2d}}) c_1(\JJ)   \right. \nn [4pt]
& \qquad  \left.  + \frac{g_1(n_-,n_+,p,q) \, g_2(n_-,n_+,p,q)}{9 n_-^3 n_+^3 (n_- + n_+)^2}\, c_1(\JJ)^2 \right\}\, ,
\end{align}
where we have defined 
\begin{align}
	\begin{split}
		f_1(n_-,n_+,p,q) & \, \equiv \,  -2 (n_-^2+n_+^2) (p+q)+(n_--n_+) (2 p^2+7 p q+4 q^2)\nn
		&\qquad +2 n_- n_+ (-n_-+n_++2 p+3 q)-q (5 p^2+5 p q+2 q^2)\, ,
	\end{split}\\[10pt]
	\begin{split}
		f_2(n_-,n_+,p,q) & \, \equiv \,  -2 (n_-^2+n_+^2) (p+q)+(n_--n_+) (2 q^2+7 p q+4 p^2)\nn
		&\qquad +2 n_- n_+ (-n_-+n_++2 q+3 p)-p (5 q^2+5 p q+2 p^2)\, ,
	\end{split}\\[10pt]
	\begin{split}
		f_3(n_-,n_+,p,q) & \, \equiv \,  -(n_--n_+) (2 p^2+p q+2 q^2)-2 n_- n_+ (p+q)\\
		&\qquad +(2 p^2-p q+2 q^2) (p+q)\, ,
	\end{split}\\[10pt]
	\begin{split}
		g_1(n_-,n_+,p,q) & \, \equiv \,  -(n_--n_+) (4 p^2+13 p q+4 q^2)+4 (n_--n_+)^2 (p+q) \nn
		& \qquad  +4 n_- n_+ (n_--n_+)+9 p q (p+q)\, ,
	\end{split}\\[10pt]
	g_2(n_-,n_+,p,q) & \, \equiv \,  (n_- - n_+)(p+q) + n_- n_+ - (p^2 + p q + q^2)\, .\nonumber
\end{align}

\subsection{$c$-extremization and superconformal R-symmetry}\label{sec:extreme}

The coefficient of $\frac{1}{2} c_1(L_i)c_1(L_j)$ in the anomaly polynomial
 $\mathcal{A}_{\mathrm{2d}}$ is precisely
$\text{Tr}\, \gamma^3 \mathcal{Q}_i \mathcal{Q}_j$, where the global symmetry $\mathcal{Q}_i$ is associated to the $U(1)$ bundle $L_i$ over $Z_4$, and $\gamma^3$ is the $d=2$ chirality operator. 
We know from $c$-extremization \cite{Benini:2012cz} that the $d=2$ superconformal $U(1)$ R-symmetry extremizes the trial function
\begin{equation}\label{ct}
c_{\mathrm{trial}} \, =\, 3\,  \text{Tr}\, \gamma^3 R_{\mathrm{trial}}^2\,,
\end{equation}
over the space of possible R-symmetries. We accordingly set
\begin{equation}\label{Rtrial}
R_{\mathrm{trial}} \, = \, R_{\mathrm{2d}} + \epsilon\,  \JJ\,,
\end{equation}
so that
\begin{equation} \label{ctriale}
c_{\mathrm{trial}}\, = \, 3 \left( \text{Tr}\, \gamma^3 R_{\mathrm{2d}}^2 +2 \epsilon\,  \text{Tr}\, \gamma^3 R_{\mathrm{2d}} \JJ + \epsilon^2\,  \text{Tr}\,  \gamma^3 \JJ^2 \right)\,.
\end{equation}
Next we can substitute for the traces using \eqref{anomaly2dfinalfinal}, and extremize over the trial 
R-charges $\epsilon$, 
$\Delta_I$, subject to the constraint $\Delta_1+\Delta_2+\Delta_3=2$. We find the 
 extremal values of these parameters to be 
\begin{equation}\label{epsilonstar}
\epsilon_* \, = \, \frac{ n_- n_+ (n_- + n_+)}{(n_--n_+) (p+q) + n_-n_+-p^2-p q-q^2}\, , \ \quad \Delta_1^* \, = \, \Delta_2^* \, = \, \Delta_3^* \, = \, \frac{2}{3} \,.
\end{equation}
The right-moving central charge, which is equal to the central charge $c$ in the large $N$ limit, is then 
given by evaluating \eqref{ct} on the superconformal R-symmetry with $\epsilon=\epsilon_*$ and $\Delta_I=\Delta_I^*$. We find 
\begin{equation} \label{cfield3charge}
c \, = \, \frac{3 p q(n_- - n_+ - p - q)}{n_- n_+ [(n_--n_+) (p+q) + n_-n_+-p^2-p q-q^2]} N^2\,.
\end{equation}
This precisely matches 
 the gravity computation \eqref{cgrav3charge}, and moreover the 
R-symmetry \eqref{Rtrial} with $\epsilon=\epsilon_*$ and $\Delta_I=\Delta_I^*$ precisely matches \eqref{mixing}!

\subsection{M5-brane anomaly polynomial}\label{sec:anomM5}

The large $N$ limit of anomaly polynomial for $N$ M5-branes is given by (see {\it e.g.}  \cite{Hosseini:2020vgl}) 
\begin{align}\label{A6d}
\mathcal{A}_{\mathrm{6d}}\, = \, \frac{1}{24}\p_2(R)N^3\, .
\end{align}
Here $R$ denotes the $SO(5)_R$ symmetry of the M5-brane theory, which geometrically is identified with the normal 
bundle to the M5-brane in spacetime, and $p_2$ denotes the second Pontryagin class. The anomaly 
polynomial \eqref{A6d} is only valid to leading order in the large $N$ limit, and more generally 
receives $O(N)$ corrections involving also Pontryagin classes  of the tangent bundle of the M5-brane. 

The supergravity solution in section \ref{sec:M5} has normal bundle twisted via 
the Cartan $U(1)\times U(1)\subset SO(5)_R$. Denoting these as two complex line bundles 
$\mathcal{N}_1$, $\mathcal{N}_2$, as in section \ref{sec:upliftM5}, 
we may write the Pontryagin class in terms of Chern classes 
so that
\begin{align}\label{A6dagain}
\mathcal{A}_{\mathrm{6d}}\, = \, \frac{1}{24}c_1(\mathcal{N}_1)^2c_1(\mathcal{N}_2)^2 N^3\, .
\end{align}
As described in section \ref{sec:upliftM5}, 
the M5-branes are wrapped on $\Sigma_g\times \Sigma$, with $\Sigma=\mathbb{WCP}^1_{[n_-,n_+]}$ 
a spindle. From the metric we saw that the normal bundles are 
\begin{align}
\mathcal{N}_1\, = \, T^*\Sigma_g\otimes L_3\, , \qquad \mathcal{N}_2\, = \, {L}_1^2\, ,
\end{align}
where ${L}_1\cong{L}_2$ because we have set the first two charges equal, $Q_1=Q_2$, 
and $T^*\Sigma_g$ is the cotangent bundle to the Riemann surface $\Sigma_g$, with total 
space being a Calabi-Yau two-fold.  The corresponding first Chern class gives minus the Euler number 
of this Riemann surface:
\begin{align}
\int_{\Sigma_g}c_1(T^*\Sigma_g)\, = \, 2(g-1)\, .
\end{align}
Since $c_1(\mathcal{N}_1) = c_1(T^*\Sigma_g) + c_1(F^{(3)})$, we may first 
integrate the anomaly polynomial \eqref{A6dagain} over $\Sigma_g$ to obtain
\be
\begin{split}
\mathcal{A}_{\mathrm{4d}}& \, = \, \int_{\Sigma_g}\mathcal{A}_{\mathrm{6d}} \, = \, \frac{N^3}{24}\cdot 2  \left(\int_{\Sigma_g}c_1(T^*\Sigma_g)\right) c_1(F^{(3)})\left[2c_1(F^{(1)})\right]^2\, \\
& \, = \,  c_1(F^{(3)})c_1(F^{(1)})^2 \cdot \frac{2}{3}(g-1) N^3\, .
\end{split}
\ee
We have thus effectively reduced to the four-dimensional anomaly polynomial 
\eqref{anomaly4d}, where we must take $F^{(1)}=F^{(2)}$ and the overall factor of $N^2/2$ should be 
replaced by $2(g-1)N^3/3$. The rest of the computation 
of the $d=2$ anomaly polynomial, obtained by integrating 
$\mathcal{A}_{\mathrm{4d}}$ over the spindle, and extracting the central charge, 
then goes through \emph{mutatis mutandis}. The final central charge 
is thus given by \eqref{cfield3charge} with $p=q$, and replacing 
an overall factor of $N^2/2$ by $2(g-1)N^3/3$. We obtain 
\begin{align}
c \, = \, \frac{4(n_--n_+-2q)q^2}{n_-n_+[n_-(n_++2q)  -q (2n_++3q)]} (g-1) N^3\, .
\end{align}
This perfectly matches the supergravity result \eqref{cM5gravity}.

\section{Discussion}\label{sec:discuss}

In this paper we have constructed a five-parameter family of 
AdS$_3\times Y_7$ solutions of type IIB supergravity, in which 
$Y_7$ is the total space of an $S^5$ fibration over a 
spindle $\Sigma=\mathbb{WCP}^1_{[n_-,n_+]}$. In 
addition to the coprime positive integers $n_->n_+$ specifying $\Sigma$, the solution 
is also characterized by positive integers $q,p$ that determine 
the twisting of the $S^5$ over $\Sigma$, together with 
the quantized five-form flux $N$ through the $S^5$. 
Setting $q=p=\frac{1}{3}(n_--n_+)$ recovers the solutions of 
\cite{Ferrero:2020laf}, and we have thus 
generalized those solutions via the addition of the 
twisting parameters $q, p$. As in \cite{Ferrero:2020laf}
we interpret these as the near-horizon limit of 
$N$ D3-branes wrapping the spindle $\Sigma$, with 
$q$ and $p$ now determining the background $U(1)^3$
Abelian R-symmetry fluxes via \eqref{introfluxes}, \eqref{introcharges}. 
As a consistency check on this interpretation, we 
have reproduced the gravity formulas for the central charge 
and superconformal R-symmetry from a dual field theory computation, 
starting with the anomaly polynomial of $\mathcal{N}=4$ SYM 
and performing an appropriate twisted compactification on $\Sigma$. 
Similarly, we have constructed a four-parameter family 
of AdS$_3\times Y_8$ solutions of $d=11$ supergravity, 
which we interpret as the near-horizon limit of $N$ 
M5-branes wrapping $\Sigma_g\times \Sigma$, 
and we have reproduced the gravitational formula for the central charge 
from a dual field theory computation.

One of the most interesting features of the solutions of \cite{Ferrero:2020laf}, 
that our solutions inherit, is that the Killing spinors on $\Sigma$ 
are not simply given by a topological twist. This is exemplified by 
the formula \eqref{introchargesum}, where the right hand side 
is not the Euler number $\chi(\Sigma)$ given in \eqref{introEuler}. 
An analogous family of $d=4$ AdS$_2\times \Sigma$  solutions
was studied in more detail in \cite{Ferrero:2020twa}. In fact these 
are near-horizon limits of full $d=4$ accelerating extremal black hole solutions, 
where it is the acceleration parameter that effectively leads to the
conical deficit singularities on the black hole horizon $\Sigma$. 
In this case we also have a UV geometry, on the conformal 
boundary of the black holes in AdS$_4$, and 
remarkably one finds that there is a topological twist 
on the copy of $\Sigma$ in the UV, but that it has 
effectively been cut in half along an equator by the acceleration horizon of the black hole! 
The spinor is a \emph{different} constant on the two halves of the spindle in the UV. 
The physical interpretation of this, in terms of wrapped branes, 
is still somewhat obscure, and it would be interesting to 
find an analogous class of black string solutions in $d=5$, $U(1)^3$ gauged supergravity that 
have our solutions as a near-horizon limit. Alternatively, 
one could directly attempt to study $\mathcal{N}=4$ SYM on 
a  rigid supersymmetric background $\R^2\times \Sigma$, with
background fluxes \eqref{introfluxes}, \eqref{introcharges}, making 
more explicit the twisting of the fields and their boundary conditions 
at the conical singularities of $\Sigma$. In particular, one might envisage 
that additional data needs to be specified at the conical singularities, in describing the 
behaviour of the fields. We note that the anomaly polynomial method 
gives the correct supergravity result, despite the presence of the conical deficit
singularities on $\Sigma$, and it would be interesting to justify this more 
carefully in such an analysis. 

Finally, in section \ref{sec:M5} we uplifted the $d=5$, 
$Q_1=Q_2$ solutions on the Maldacena-N\'u\~nez solution, 
but we also commented that one can uplift on 
other internal six-manifolds 
$N_6$ to obtain solutions to $d=11$ supergravity. Such solutions  \cite{Gaiotto:2009gz}
describe M5-branes wrapped on a Riemann surface, in general with punctures, 
which are then further wrapped on the spindle $\Sigma$ to obtain 
$d=2$, $(0,2)$ SCFTs. It would be interesting to investigate these solutions 
in more detail, in particular computing the central charge 
in gravity and via the anomaly polynomial for the theory on $N$ M5-branes. 
We leave this, together with other interesting generalizations, for 
future work.

\subsection*{Acknowledgments}
We thank Dario Martelli for helpful discussions. The work of JFS  was supported in part by STFC grant ST/T000864/1.

\bibliographystyle{utphys}
\bibliography{helical}{}
\end{document}